\documentclass[twocolumn,prb,showpacs,preprintnumbers,amsmath,amssymb,superscriptaddress]{revtex4-1}
\usepackage{graphicx}
\usepackage{setspace}

\usepackage{color}
\usepackage{ulem} 

\usepackage{amsmath,amsthm,amssymb}
\usepackage{mathrsfs}

\usepackage[%dvipdfmx,
colorlinks=true, citecolor=blue, urlcolor=blue, linkcolor=blue,
setpagesize=false, bookmarks=false
]{hyperref}

\begin{document}

\title{Excitonic condensation reflecting electronic states \\
  in two-band Penrose-Hubbard model}% Force line breaks with \\

\author{Ken Inayoshi}
\email{k-inayoshi@stat.phys.titech.ac.jp}
\author{Yuta Murakami}%
\author{Akihisa Koga}%
 
\affiliation{%
Department of Physics, Tokyo Institute of Technology, Meguro, Tokyo 152-8551, Japan
}%

\date{\today}% It is always \today, today,
             %  but any date may be explicitly specified

\begin{abstract}
  We study the excitonic insulating (EI) phase in the two-band Hubbard models
  on the Penrose tiling.
  Performing the real-space mean-field calculations systematically,
  we obtain the ground state phase diagrams for the vertex and center models.
  We find that, in some regimes, the stable EI phase is induced
  by small interband interactions.
  We argue that this originates from the electron-hole pairing for  the completely or nearly degenerate states,
  which are characteristic of the Penrose tiling. 
  We also study spatial distribution of the order parameter,
  mapping it to the perpendicular space.
\end{abstract}

%\keywords{Suggested keywords}%Use showkeys class option if keyword
                              %display desired
\maketitle

%\tableofcontents

\section{INTRODUCTION}
Quasicrystals (QC) are exotic systems
which have no translational symmetry but have long-range order and
special rotational symmetry.
Since the first discovery of
a QC material 
\cite{PhysRevLett.53.1951,PhysRevLett.53.2477},
intensive efforts have been done to study various properties inherent in QC materials
such as the lattice structure and electric characteristics~\cite{PhysRevLett.55.2883,Tsai_1987,Tsai_1988,199098,RevModPhys.65.213,Tsai2000}.
One of the important examples is the Au-Al-Yb arroy~\cite{Ishimasa},
where quantum critical behavior appears in the quasicrystal and
heavy fermion behavior appears in the approximant~\cite{Deguchi}. Recently, the superconductivity has been observed in Al-Zn-Mg quasicrystalline
alloys~\cite{qcsupercondexp}.
These experiments stimulate %the possibility of various ordered phases in QCs and provoked
theoretical studies on
electron correlations~\cite{Watanabe,Takemori,Takemura,Andrade,Shinzaki,Otsuki} and ordered phases
in the quasiperiodic systems~\cite{PhysRevB.95.024509,PhysRevB.96.214402,PhysRevResearch.1.022002}.
On the other hand, in research on electronic structures,
there has been a question whether or not semiconducting QC materials exist
since semiconductors and insulators are never synthesized up to now.
Recently,  it has been reported that the cubic quasicrystalline approximant
Al-Si-Ru~\cite{PhysRevMaterials.3.061601} has a semiconducting band structure,
where the band gap appears between conduction and valence bands.
Therefore, it is important and interesting to study
%the potential ordered states and their properties
electron correlations in the quasiperiodic systems with multiband.

One of the important topics in multiband correlated electron systems is the excitonic insulating (EI) phase.
This phase was proposed more than 50 years ago as the spontaneous condensation of electron-hole pairs
in multiband systems induced by the interband Coulomb interaction,
which is theoretically analogous to the superconducting
phase~\cite{jerome1967,kohn1967,keldysh1968,halperin1968,halperin1968RMP}.
Although the EI phase in real materials has been elusive for decades,
this phase has recently been attracting interest due to signatures of the EI phase
in several candidate materials such as
$\rm{Te_{2}NiSe_{5}}$~\cite{PhysRevLett.103.026402,Wakisaka2012} and
$1T$-TiSe$_{2}$~\cite{PhysRevLett.99.146403,PhysRevLett.106.106404}.
Therefore, it is instructive to discuss the possibility of the EI state in the quasiperiodic systems.

In this paper, we study the two-band Hubbard model
%on the two-dimensional Penrose tiling
to reveal how the EI phase emerges in the quasiperiodic systems and
clarify its characteristic features.
We here treat the Penrose tiling~\cite{penroselattice} as a prototypical structure
for theoretical studies on the quasiperiodic
systems~\cite{Choy_1985,Kohmoto,Kohmoto2,Sutherland,Tsunetsugu_1986,Kumar_1986,Odagaki,Kohmoto_1987,PhysRevB.38.1621,PhysRevB.37.2797,Tsunetsugu1,Tsunetsugu2,Takemori,Takemura,Shinzaki,PhysRevB.95.024509,PhysRevB.96.214402,PhysRevResearch.1.022002}.
Applying the real-space mean-field (MF) theory to the Hubbard model on the quasiperiodic
lattice~\cite{Jagannathan_1997,PhysRevB.96.214402},
we study the stability of the EI phase and determine the ground state phase diagram.
We find that the EI phase emerges even with small interband interaction in some cases.
It is clarified that
this originates from the existence of the degenerate (confined states
and string states~\cite{PhysRevB.38.1621,PhysRevB.37.2797}) or
nearly degenerate states characteristic of the Penrose tiling.
Examining the spatial distribution of the order parameters in the real and perpendicular spaces,
we also discuss the crossover behavior in the EI phase. 

%%%%%%%%%%%%%%%%%%%%%%%%%%%%%%%%%%%%%%%%%%%%%%%%%%%%%%%
\section{Model and Method} 
\label{sec:sec2}
%%%%%%%%%%%%%%%%%%%%%%%%%%%%%%%%%%%%%%%%%%%%%%%%%%%%%%%

In this work, we consider the two-band Hubbard models to
discuss the stability of the EI state at zero temperature.
The Hamiltonian is given as 
\small
\begin{align}
\hat{H}&=-\sum_{\langle i,j \rangle \sigma}(t\hat{c}_{i\sigma}^{\dagger}\hat{c}_{j\sigma}-t\hat{f}_{i\sigma}^{\dagger}\hat{f}_{j\sigma}) 
 +\frac{D}{2}\sum_{i\sigma}(\hat{n}_{ci\sigma}-\hat{n}_{fi\sigma}) \nonumber \\ 
 &-\mu\sum_{i\sigma}(\hat{n}_{fi\sigma} +\hat{n}_{ci\sigma}) \nonumber \label{hamiltonian}\\
&+U\sum_{i}(\hat{n}_{ci\uparrow}\hat{n}_{ci\downarrow}+\hat{n}_{fi\uparrow}\hat{n}_{fi\downarrow})+V\sum_{i\sigma\sigma^{'}}\hat{n}_{ci\sigma}\hat{n}_{fi\sigma'},
\end{align}
\normalsize
where $\hat{f}^\dagger_{i\sigma}$ ($\hat{c}^\dagger_{i\sigma}$) is
a creation operator of the electron at site $i$
with spin $\sigma\in \{ \uparrow, \downarrow \}$
in the f-band (the c-band),
$\hat{n}_{ci\sigma}=\hat{c}^\dagger_{i\sigma}\hat{c}_{i\sigma}$
and $\hat{n}_{fi\sigma}=\hat{f}^\dagger_{i\sigma}\hat{f}_{i\sigma}$.
$t$ is the hopping integral between nearest neighbor sites, $D$ is the difference between the energy levels of two bands,
$\mu$ is the chemical potential, 
$U(>0)$ is the intraband on-site interaction, and
$V(>0)$ is the interband interaction.
This Hamiltonian is invariant under the transformations
$c_{i\sigma}\leftrightarrow f_{i\sigma}^\dag$
when the chemical potential is fixed as $\mu=\frac{U}{2}+V$.
In the case, $\langle \hat{f}_{i\sigma}^{\dagger}\hat{f}_{i\sigma} \rangle+
\langle \hat{c}_{i\sigma}^{\dagger}\hat{c}_{i\sigma} \rangle=1$ is always satisfied at each site,
and we focus on such a case in the following.

%\subsection{Real-Space Mean-Field Theory}
Here, we use the real-space MF theory
to determine the ground state phase diagram of
the two-band Hubbard model.
To focus on the stability of the EI state,
our discussions are restricted to be paramagnetic.
Then, we introduce three MF parameters as 
\begin{eqnarray}
n_{fi}&=&\langle\hat{f}_{i\sigma}^{\dagger}\hat{f}_{i\sigma}\rangle, \\
n_{ci}&=&\langle\hat{c}_{i\sigma}^{\dagger}\hat{c}_{i\sigma}\rangle, \\
\Delta_{i}&=&\langle\hat{c}_{i\sigma}^{\dagger}\hat{f}_{i\sigma}\rangle,\label{orderparameter}
\end{eqnarray}
where $\langle \cdots \rangle$ is the grand canonical average and
$\Delta_i$ is the order parameter of the EI state at site $i$.
Then, the MF Hamiltonian is obtained as,
\begin{align} \label{eq:MF}
\hat{H}_{\rm MF}&=\sum_{\langle i,j \rangle \sigma}(t\hat{f}_{i\sigma}^{\dagger}\hat{f}_{j\sigma}-t\hat{c}_{i\sigma}^{\dagger}\hat{c}_{j\sigma}) +\frac{D}{2}\sum_{i\sigma}(\hat{n}_{ci\sigma}-\hat{n}_{fi\sigma}) \notag \\
&+(2V-U)\sum_{i\sigma} [(n_{ci}-\frac{1}{2})\hat{n}_{fi\sigma}+(n_{fi}-\frac{1}{2})\hat{n}_{ci\sigma}]\notag\\
&-V\sum_{i\sigma}(\Delta_{i}\hat{f}_{i\sigma}^{\dagger}\hat{c}_{i\sigma}+h.c.),
\end{align}
where the third (fourth) term represents Hartree (Fock) contributions.
In our study, we consider the system with $U/V=2$, for simplicity,
where the Hartree term vanishes.

\begin{figure}[t]
\centering
\includegraphics[width=\linewidth]{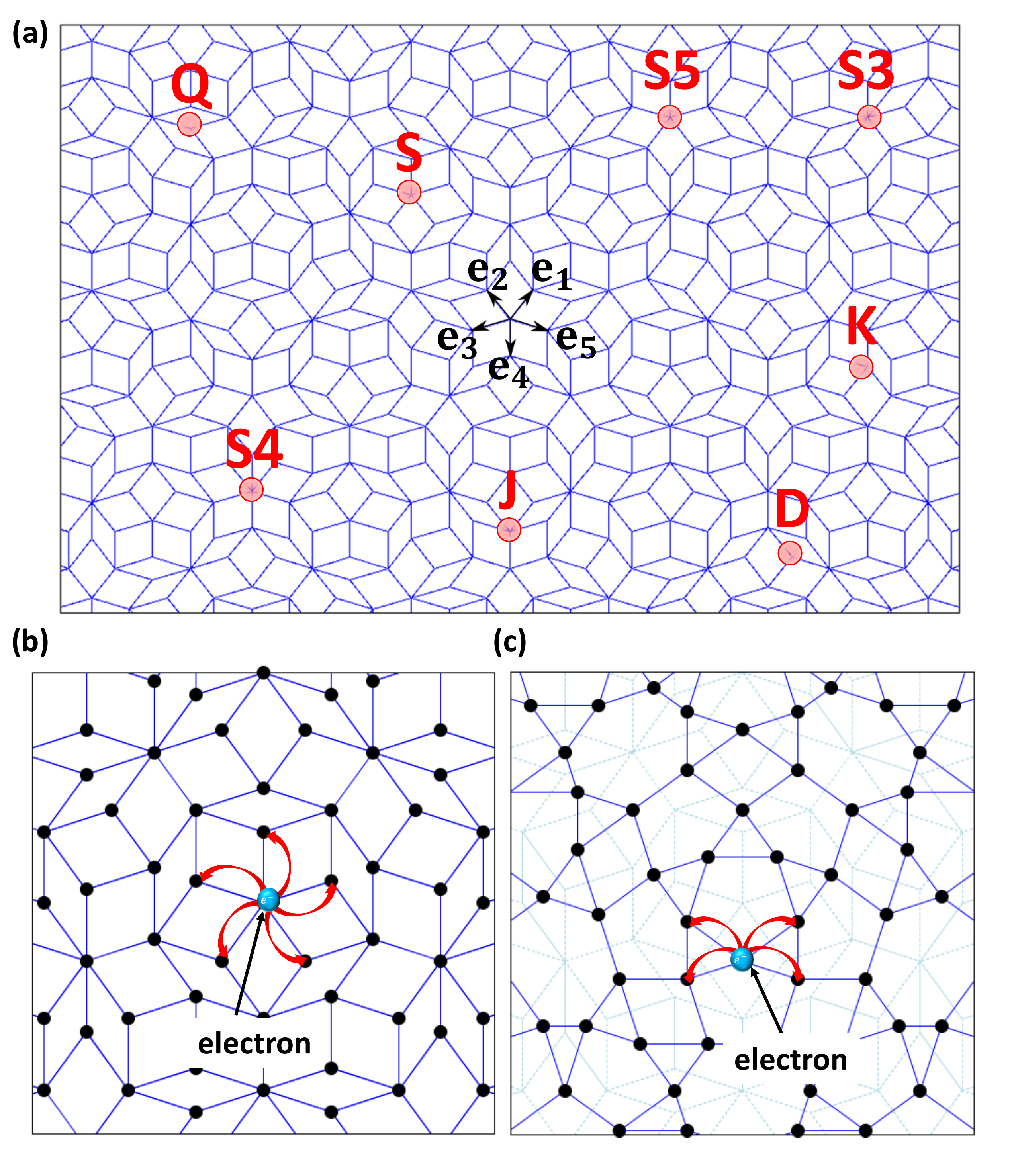}
\caption{
  (Color online)
  (a) Penrose tiling and eight types of vertices~\cite{Bruijn1,Bruijn2}.
  $e_{1},\cdots,e_{5}$ are projection of the translational vectors in five dimensions,
  ${\bf n}=(1,0,0,0,0),\cdots,(0,0,0,0,1)$. 
  (b)(c)Schematic pictures of (b) vertex and (c) center models. 
  }
\label{VMandCM}
\end{figure}

We consider the EI state in
the two-band system on the quasiperiodic lattice.
To this end, we treat the Penrose tiling as a simple lattice structure,
which is shown in Fig.~\ref{VMandCM}(a).
The lattice is composed of the fat and skinny rhombuses
and includes eight kinds of vertices~\cite{Bruijn1,Bruijn2}.
It is known that there are two ways to construct the tight-binding models on the Penrose tiling,
so-called, the vertex and center models, as shown in Figs.~\ref{VMandCM}(b) and (c).
In the vertex model, lattice sites are located
on the vertices of the Penrose tiling and
electrons hop along the edges of the rhombus,
as shown in Fig.~\ref{VMandCM}(b).
The center model is deviated from the vertex model, where
electrons are placed at the center of the rhombuses and
hopping integral is finite only between the centers of neighboring rhombuses
with shared face, as shown in Fig.~\ref{VMandCM}(c).
In the calculation, we construct the lattices with finite $N$
which are generated by the dilatation operation called {\it deflation}~\cite{Bruijn2}.
%%%%%%%%%%%%%%%%%%%%%%%%%%%%%%%
%%%%%%%%%%%%%%%%%%%%%%%%%%%%%%%

When $\Delta_i=0$, the normal state is realized and
the effect of electron correlations is irrelevant in the MF Hamiltonian eq. (\ref{eq:MF}). 
In this case, the system is described by
two distinct single-band noninteracting models
with the energy shifts $\pm D/2$.
Therefore, the DOS structure for the single-band model $\hat{h}=-t\sum_{ij} \hat{a}^\dag_i \hat{a}_j$
should be useful to understand the ground state properties in the weak-coupling regime.
%%%%%%%%%%%%%%%%%%%%%%%%%%%%%%%%
\begin{figure}[t]
\centering
\includegraphics[width=\linewidth]{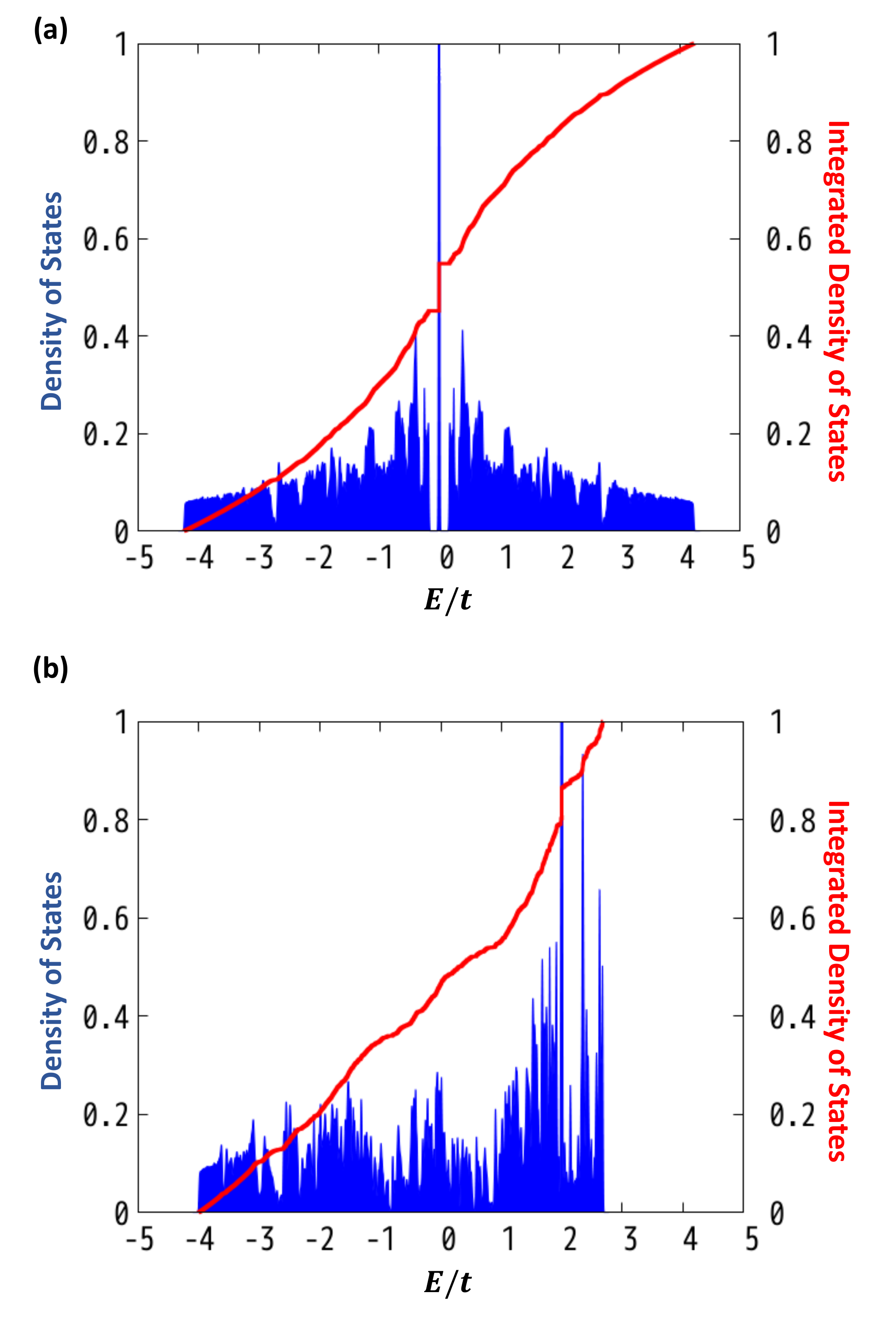}
\caption{
  (Color online) Densities of states and integrated density of states for the single-band Hamiltonian of
  (a) vertex and (b) center models on the Penrose tiling,
  which are obtained by full diagonalizations for the systems
  with $N=28901$ and $N=28270$, respectively.
}
\label{DOS}
\end{figure}
%%%%%%%%%%%%%%%%%%%%%%%%%%%%%%%%%
Figure~\ref{DOS} shows the DOS for the single-band tight-binding Hamiltonians
for the vertex and center models.
The vertex model is bipartite, leading to a symmetric energy spectrum of electrons
with $W_-\simeq-4.2t$, $W_+\simeq4.2t$, and $W=W_+-W_-\simeq8.4t$,
where $W_\pm$ is the upper and lower bound of the DOS and $W$ is the bandwidth.
There is a delta function peak at $E=0$ in the DOS,
which originates from the macroscopic number of degenerate states.
These states are known as the confined states,
whose wave function is spatially confined to
a finite range~\cite{Kohmoto,PhysRevB.38.1621,PhysRevB.96.214402}.
In addition, we find a finite energy gap between the delta function peak and
the continuous spectrum in the DOS.
The gap size has been obtained as $\Delta_G \sim 0.17t$~\cite{Kohmoto2,PhysRevB.96.214402}.
In contrast to the vertex model, the center model is no longer bipartite,
as shown in Fig.~\ref{VMandCM}(c).
Then, an asymmetric energy spectrum appears due to geometrical frustration,
$W_-\simeq-4t$, $W_+\simeq2.7t$ and $W\simeq6.7t$, 
as shown in Fig.~\ref{DOS}(b).
In the center model, we could not find the gap in the DOS, but
there exist not only the delta function peak at $E=2t$,
but also a sharp peak at $E\simeq2.35t$.
The former again is associated with the macroscopic number of degenerate states,
and they consist of the confined states and the so-called string states~\cite{PhysRevB.37.2797}.
Therefore, interesting ground state properties inherent in the center model
are also expected.

Now we turn to the two-band models with $\Delta_i=0$ on the Penrose tiling.
When $D>2|W_-|$, all energy levels of the c-band (f-band) are located above (below)
the Fermi level, and the insulating state is realized.
In the case $-2W_+<D<2|W_-|$,
the Fermi level is located inside of both c- and f-bands,
and the metallic state is realized, except for $|D|<2\Delta_G$ in the vertex model.
When $D<-2W_+$, the c-band is located below the Fermi level and the f-band
is above it, which leads to the insulating state.
In the following, we discuss how the introduction of the interband interactions
affects ground state properties, and, in particular,
reveal the characteristic features of EI states on the Penrose tiling.

%%%%%%%%%%%%%%%%%%%%%%%%%%%%%%%%%%%%%%%%%%%%%
\section{Results}\label{sec:sec3}
%%%%%%%%%%%%%%%%%%%%%%%%%%%%%%%%%%%%%%%%%%%%%

%\subsection{The Vertex model on the Penrose tiling}

\begin{figure}[t]
\centering
\includegraphics[width=0.9\linewidth]{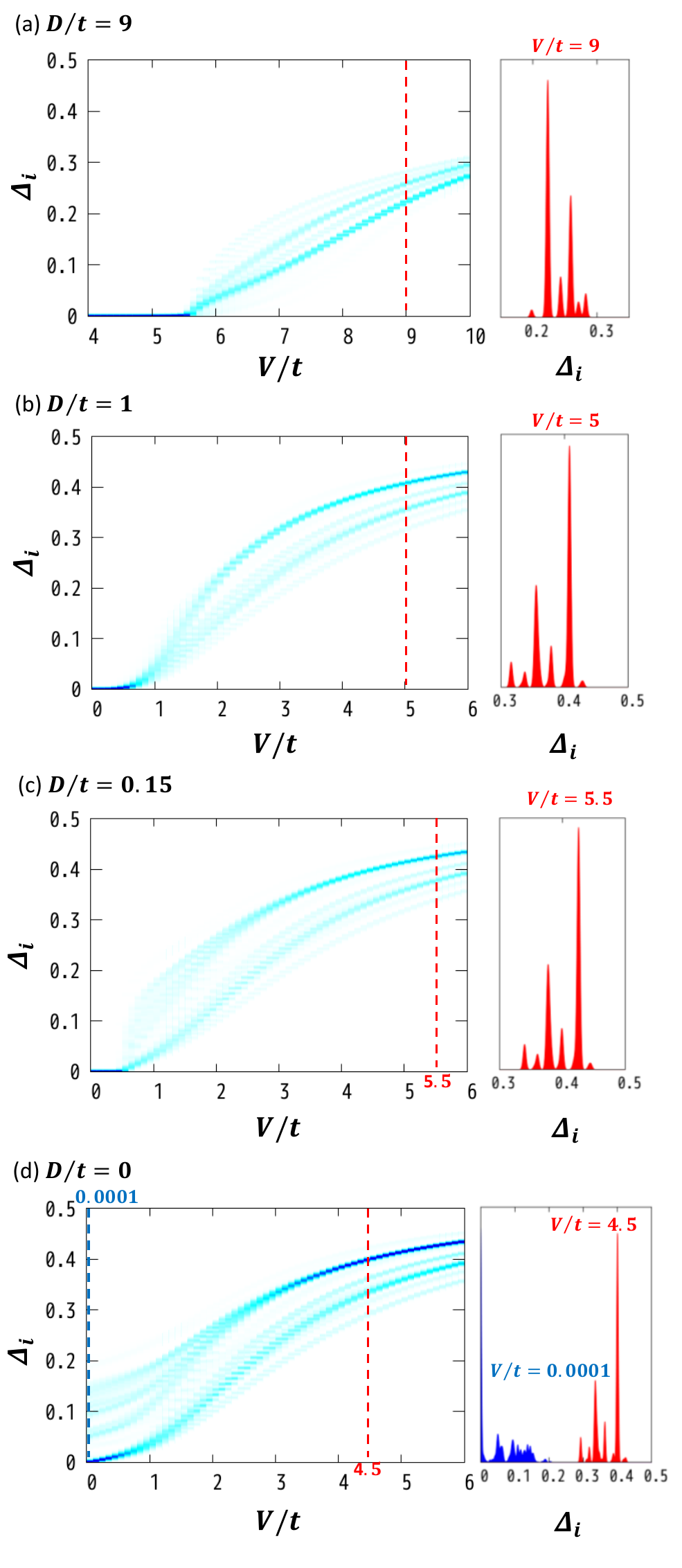}
\caption{
  (Color online) Density plot of the EI order parameter in the vertex model with $N=11006$
  for (a) $D/t=9$, (b) $D/t=1$, (c) $D/t=0.15$ and (d) $D/t=0$.
  The right panels show the cross section for indicated $V/t$.}
\label{VMDensity}
\end{figure}

In this section, we discuss how the EI state is realized
in the two-band Hubbard model on the Penrose tiling.
First, we focus on the vertex model, as shown in Fig.~\ref{VMandCM}(b).
The tight-binding model is invariant under the particle-hole transformation, and 
the symmetric DOS appears for each band,
as shown in Fig.~\ref{DOS}(a). % with $W=8.4t$.
Therefore, our discussions can be restricted to the case $D\ge 0$.
%The two-band model in the noninteracting case is
%metallic when $|D| < W?$ since the Fermi level is located insite of both $c$ and $f$ bands.
%On the other hand, in the case with $|D|>W$, the Fermi level is located
%between the c and f band, where the insulating state is realized.
We now discuss how the interband interaction $V$ induces the EI states.
In the quasiperiodic system, lattice sites are, in general, different from each other,
which leads to the inhomogeneous spatial distribution of the physical quantities.
Therefore, we show in Fig.~\ref{VMDensity} the density plot of the local EI order parameters
in the two-band system with $N=11006$.
%%%%%%%%%%%%%%%%%%%%%%%%%%%%%%%%
%%%%%%%%%%%%%%%%%%%%%%%%%%%%%%%%%
When $D/t=9$, the noninteracting system ($V=0$) belongs to the band insulating state
with the energy gap $\sim D-2|W_{-}|$.
This state is stable in the case $V<V_c(\sim 5.5t)$,
where no EI order parameter appears.
Beyond the critical value $V_c$, the order parameters for all sites
are simultaneously induced, and  the second-order quantum phase transition
occurs to the EI state.
This is consistent with the results in the conventional phase transition
in the uniform lattices~\cite{sugimoto2016}.
Further increase of the interaction monotonically increases the order parameters,
where some peak structures appear in the density of the order parameters,
as shown in Fig.~\ref{VMDensity}(a).
This means that the site dependence appears in the local quantities.
In the strong coupling case, the order parameters are almost classified into the five groups (six groups if we include the edge of the lattice),
which are related to the coordination number
[see the right panel of Fig.~\ref{VMDensity}(a)].
This indicates that the correlation effect only depends on the coordination number in this regime.

When $2\Delta_G<D<2|W_{-}|$ and $V=0$,
the metallic state is realized with the finite DOS at the Fermi level $(0<\rho(E_F)<\infty)$.
Switching on the interband interactions,
the order parameters are slowly induced, as shown in Fig.~\ref{VMDensity}(b).
This behavior is common to that in the conventional ordered states in the weak coupling limit
such as the antiferromagnetical ordered state
in the half-filled Hubbard model on the bipartite lattice.
Therefore, we expect $\Delta\sim \exp(-c/V)$ where $c$ is a constant
although it is hard to confirm in our finite size calculation.
Further increase of the interaction strength beyond the crossover region $V/t\sim 1$
rapidly increases the order parameter at each site.
Then, we clearly find site-dependent order parameters.

When $0<D<2\Delta_G$ and $V=0$, the Fermi level is located in the gap
between the delta function peak and continuous spectrum in both c and f-band,
and thereby the noninteracting system is insulating.
In the case, the interband interaction yields the quantum phase transition
to the EI state,
which is similar to the case with $D>2|W_{-}|$ [see Fig.~\ref{VMDensity}(c)].
Note that when the system is close to $D=0$,
the confined states forming the delta function peak in the noninteracting DOS
should play a major role for the quantum phase transition.
Around the phase transition point ($V\lesssim V_C$),
wave functions for these states are no longer eigenstates,
which forms the band structure with a finite width due to the finite interband interaction.
Although this effect cannot be taken into account in the MF treatment correctly,
the second order phase transition should occur in the $V\neq 0$ case.
When $D=V=0$, the Fermi level is located at the energy level
for the macroscopically degenerate states in both c and f-bands.
Therefore, the introduction of the interband interaction immediately forms
electron-hole pairs and lifts the ground state degeneracy,
yielding the excitonic order parameters.
In fact, finite $\Delta_i$ appear even at the infinitesimal $V/t$
although most of the sites do not have order parameters, as shown in Fig.~\ref{VMDensity}(d).
This originates from the fact that the confined states at $E=0$ have a
finite amplitude only at certain lattice sites,
which is essentially the same as that
in the antiferromagnetical ordered state in the single band Hubbard
model~\cite{PhysRevB.96.214402}.
With increasing the interband interactions, the order parameters monotonically increase.
At last, the EI properties are almost described by the local site properties,
where the distribution of $\Delta_i$ is classified by the coordination number.

%%%%%%%%%%%%%%%%%%%%%%%%%%%%%%%%%%%%%%%%%%%%%%%%%%%%%
\begin{figure}[t]
\centering
\includegraphics[width=\linewidth]{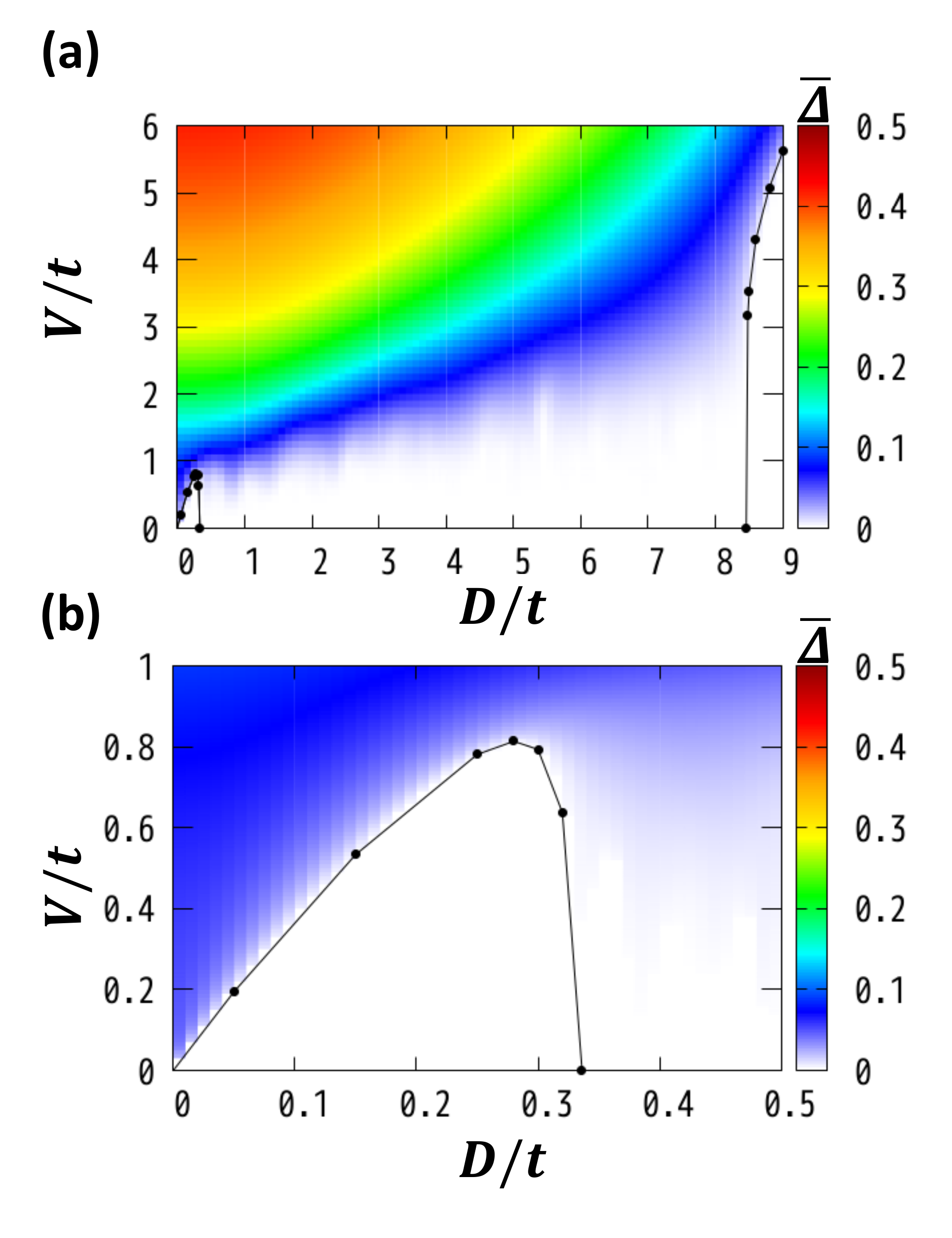}
\caption{
  (Color online)
  (a) The order parameter of the EI phase $\bar{\Delta}$ for the vertex model with $N=4181$.
  (b) The magnified image around $(D/t,V/t)=(0,0)$ in (a).
  Solid circles represent second-order quantum phase transition points.
}
\label{vertexPD}
\end{figure}
%%%%%%%%%%%%%%%%%%%%%%%%%%%%%%%%%%%%%%%%%%%%%%%%%%%%%%

By performing similar calculations for different parameters,
we obtain the phase diagram of the vertex model,
based on the spatial average of the order parameters $\bar{\Delta}(=\sum_{i=1}^{N} \Delta_i/N)$.
The density plot of $\bar{\Delta}$ in the system with $N=4181$
is shown in Fig.~\ref{vertexPD}(a).
When $V/t=0$, the metallic phase is realized in the case of $2\Delta_G<D\leq 2|W_{-}|$
and the band insulating state is in the case with $0<D<2\Delta_G$ and $2|W_{-}|<D$.
The insulating state is stable against the small interband interaction.
The EI state appears beyond a certain critical value $V_c$,
as discussed above.
With decreasing $D(>2|W_{-}|)$, the critical value $V_c$ decreases and reaches zero at $D=2|W_{-}|$.
Therefore, the EI state is realized when $2\Delta_G<D<2|W_{-}|$ and $V/t>0$ %or $V/t\neq 0$
although the magnitude of the order parameters is exponentially small
in the small $V$ region.
By contrast, finite magnitude appears around $D/t\sim 0$ and $V/t\sim 0$,
as shown clearly in Fig.~\ref{vertexPD}(b).
This phenomenon is inherent in the vertex model, which originates from
the existence of the macroscopically degenerate states at $E=0$, as discussed above.

%%%%%%%%%%%%%%%%%%%%%%%%%%%%%%%%%%%%%%%%%%%%%%%%%%%%%%%
%\subsection{The center model of the Penrose lattice}
%%%%%%%%%%%%%%%%%%%%%%%%%%%%%%%%%%%%%%%%%%%%%%%%%%%%%%%
Now, we turn to the center model [see Fig.~\ref{VMandCM}(c)] to discuss the stability of the EI state.
The interesting point for the center model is that
the noninteracting system has an asymmetric DOS for each band,
where the delta-function peak is located away from its center [see Fig.~\ref{DOS}(b)].
This allows us to discuss the EI state in the quasiperiodic lattice more generally.
%%%%%%%%%%%%%%%%%%%%%%%%%%%%%%%%
\begin{figure}[t]
\centering
\includegraphics[width=\linewidth]{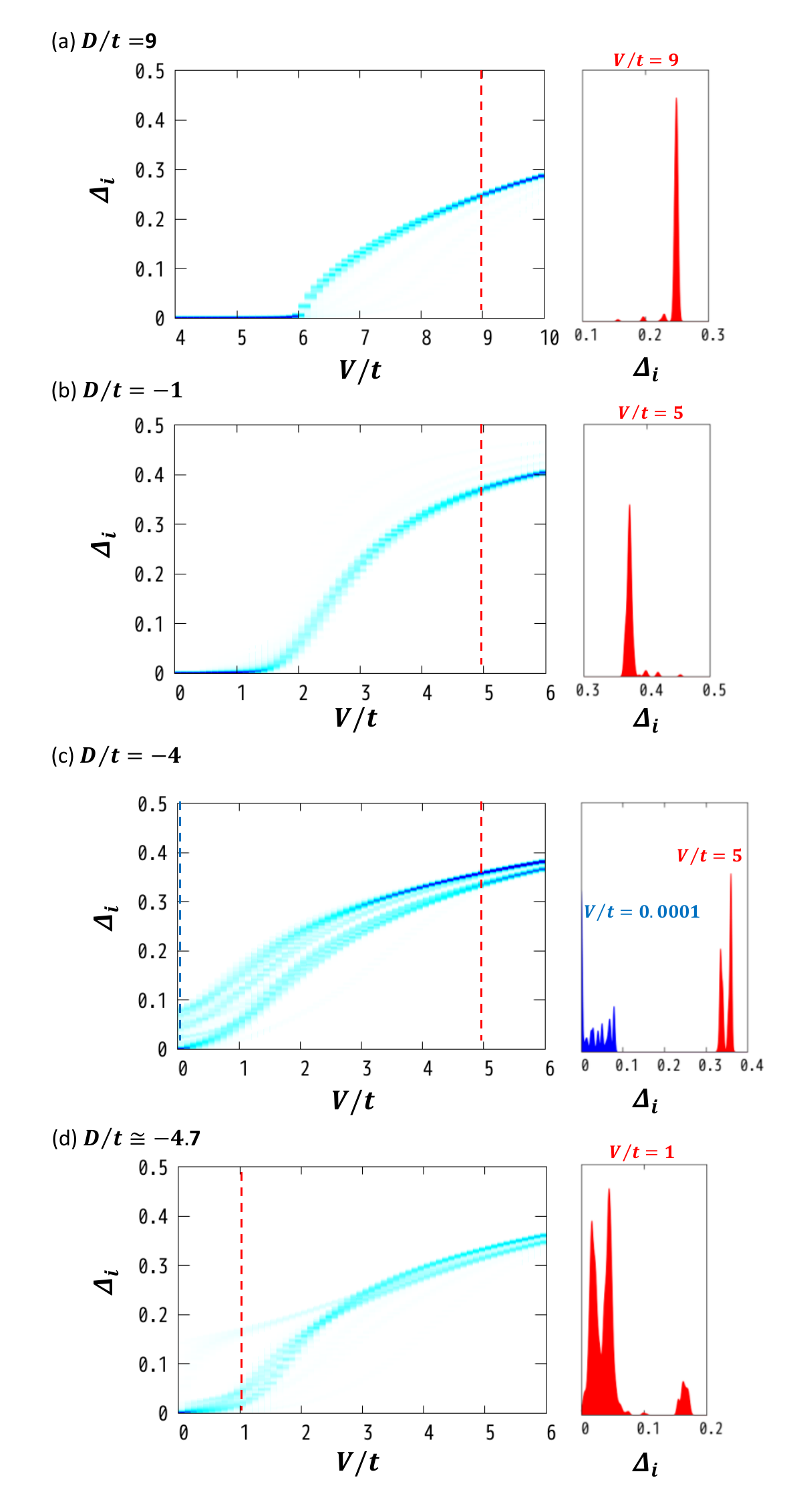}
\caption{
  (Color online) Density plots of the EI order parameter in the center model with $N=10640$
  for (a) $D/t=9$, (b) $D/t=-1$, (c) $D/t=-4$, and (d) $D/t=-4.69085$.
  Right panels show the cross section for indicated $V/t$. }
\label{CMDensity}
\end{figure}
%%%%%%%%%%%%%%%%%%%%%%%%%%%%%%%%%
Figure~\ref{CMDensity} shows the density plot of the EI order parameters
as a function of the interband interaction.
When the $D/t=9$ and $V=0$, the system is in the insulating state.
The large interband interactions drive the system to the EI state.
In contrast to the vertex model, we could not find the clear site-dependence
in the order parameters beyond the critical interactions.
This should be originated from the fact that the coordination number in the center model
is four except for the edge of the lattice.
Therefore, in the case, it is hard to see ground state properties inherent in the center model,
which are essentially the same
as those in the two-band model on the square lattice.
Similar correlation effects are also found in the case with $D/t=-1$, where the system is metallic at $V=0$ [see Fig.~\ref{CMDensity}(b)].

When $D/t=-4$, different behavior appears, as shown in Fig.~\ref{CMDensity}(c).
In the weak coupling regime, we find some peak structures
in the density plot.
This means the site-dependence in the order parameters, in contrast to the above cases.
When $V=0$,
the Fermi level is located at the energy level for the macroscopically degenerate states in each band,
which is similar to the case with $D=0$ for the vertex model.
Therefore, the spatial dependence of the order parameter reflects these states.
%Therefore, in the small $V$ region, the spatial dependence in the order parameters
%appears characteristic of the macroscopic degenerate states.
When $D/t\simeq -4.7$, site-dependence in the order parameters is also found
slightly away from $V=0$, as shown in Fig.~\ref{CMDensity}(d).
We will discuss the spatial distribution of the EI order parameters
in these two cases in detail.

%%%%%%%%%%%%%%%%%%%%%%%%%%%%%%%%%%%%%%%%%%%%%%%%%
\begin{figure}[t]
\centering
\includegraphics[width=\linewidth]{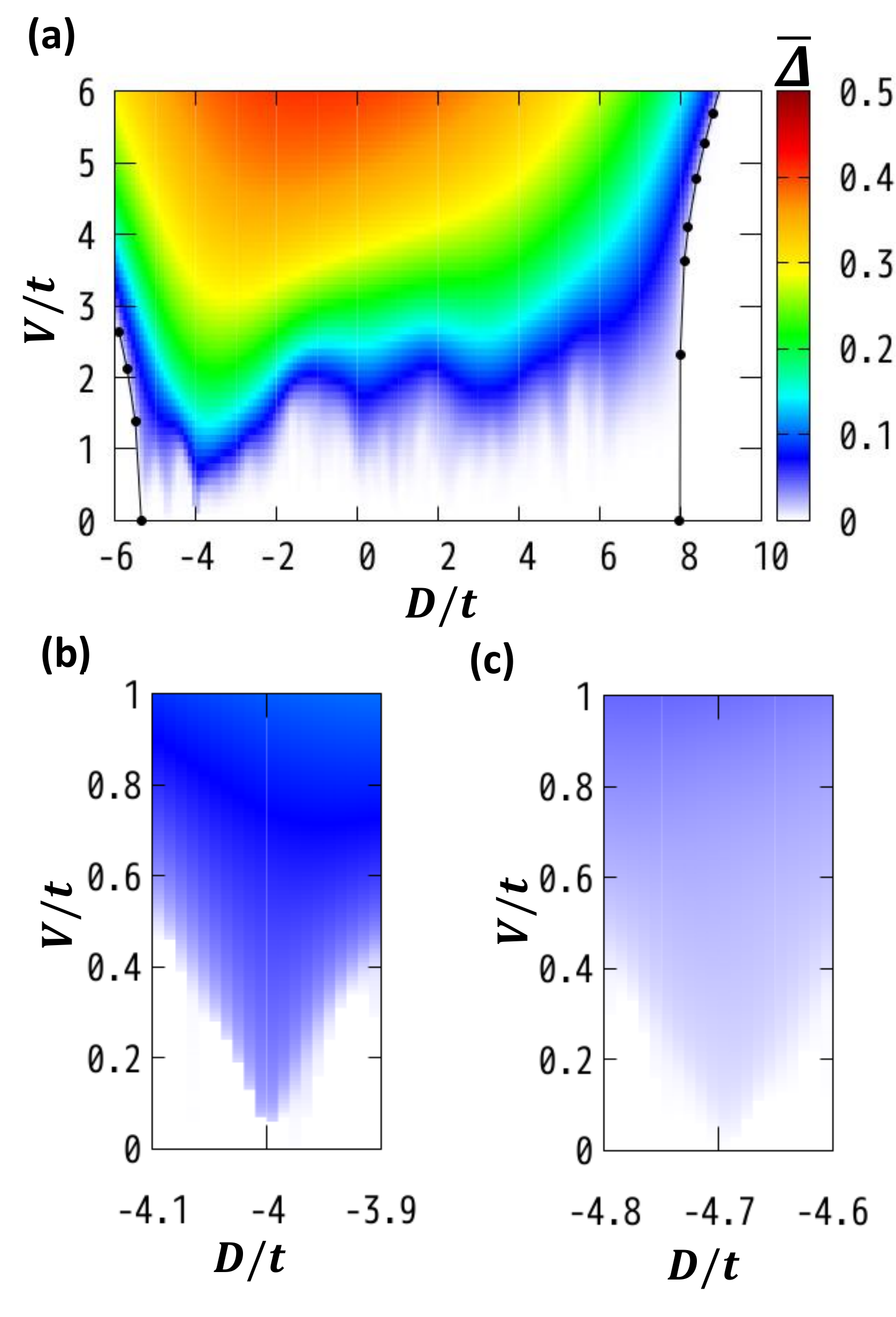}
\caption{
  (Color online) (a) The order parameter of the EI phase $\bar{\Delta}$
  as a function of $D$ and $V$ for the center model with $N=3975$.
  Solid circles represent the second-order quantum phase transition points.
  (b) and (c) show the magnified images around $(D/t,V/t)=(-4,0.0)$ and $(-4.7,0.0)$,
  respectively. }
\label{centerPD}
\end{figure}
%%%%%%%%%%%%%%%%%%%%%%%%%%%%%%%%%%%%%%%%%%%%%%%%
In Fig.~\ref{centerPD}(a), we show the average of the order parameters $\bar{\Delta}$ of
the center model.
The phase diagram is qualitatively the same as that for the vertex model.
A remarkable point is that the large intensity appears
around $D/t=-4$ and $D/t=-4.7$ at the small interacting regions
[see  Figs.~\ref{centerPD}(b) and (c)].

Here, we discuss the spatial pattern of the order parameter 
in these cases.
%%%%%%%%%%%%%%%%%%%%%%%%%%%%%%%%%%%%%%%%%%55
\begin{figure}[t]
\centering
\includegraphics[width=\linewidth]{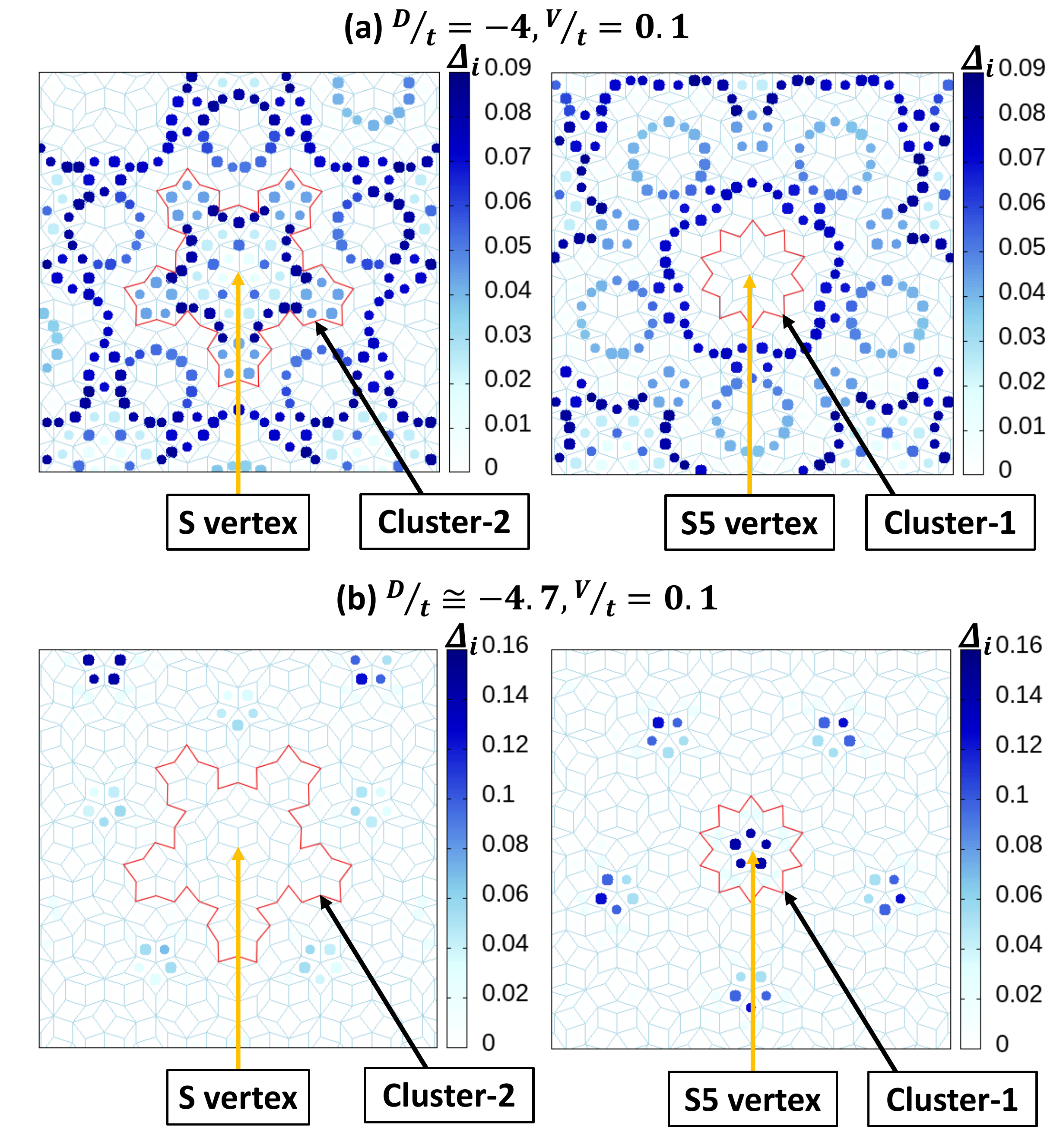}
\caption{
  (Color online) Spatial distributions of the EI order parameters  
  when (a) $(D/t,V/t)=(-4,0.1)$ and (b) $(D/t,V/t)\simeq(-4.7,0.1)$ in the center model with $N=28270$.
  Left and right panels show the results around {\it cluster-2} and {\it cluster-1},
  which are shown with the red lines.
}
\label{kousatu3}
\end{figure}
%%%%%%%%%%%%%%%%%%%%%%%%%%%%%%%%%%%%%%%%%%%%
Figure~\ref{kousatu3} shows the spatial distribution of
the local order parameters at $(D/t,V/t)=(-4, 0.1)$  and $(-4.7, 0.1)$.
Now, we focus on two regions around
{\it cluster-1} and {\it cluster-2},
which are defined in Ref.~\cite{PhysRevB.96.214402} and
are shown as the red lines in Fig.~\ref{kousatu3}.
It is known that the cluster-1 (cluster-2) is composed of the 20 (70) rhombuses and
one S5 (S) vertex is located at its center.
Note that all S5 (S) vertices are not located at the center of cluster-1 (cluster-2).
When $(D/t,V/t)=(-4,0.1)$,
distinct behavior appears in the spatial distribution of the local order parameters
in these clusters.
Namely, in the cluster-1, no order parameter appears, while
finite amplitudes appear in the cluster-2, as shown in Fig.~\ref{kousatu3}(a).
These results can be explained by the following.
It is known that in the cluster-1, there exist no confined states.
On the other hand, in the cluster-2, there exist some confined states, i.e.
so-called A1, A2, and B states, and a string state~\cite{PhysRevB.37.2797},
which leads to the star-like island with large amplitudes.
Thus, the spatial pattern of the order parameters reflects
wave functions of the macroscopically degenerate states.

On the other hand, when $(D/t,V/t)\simeq(-4.7,0.1)$, different pattern appears,
as shown in Fig.~\ref{kousatu3}(b).
We clearly find the high density only around S5 vertices
in the distribution of the EI order parameters,
in particular, around the center of the cluster-1.
On the other hand, the order parameters are almost zero inside of the cluster-2,
and this is consistent with the fact that the amplitude of the wave functions
for $E/t\simeq 2.35$ appears only in the vicinity of S5 vertices.
Thus, the spatial distribution of the EI order parameters is quite different from
that in the case with $D/t=-4$ discussed above.

\begin{figure}[t]
\centering
\includegraphics[width=\linewidth]{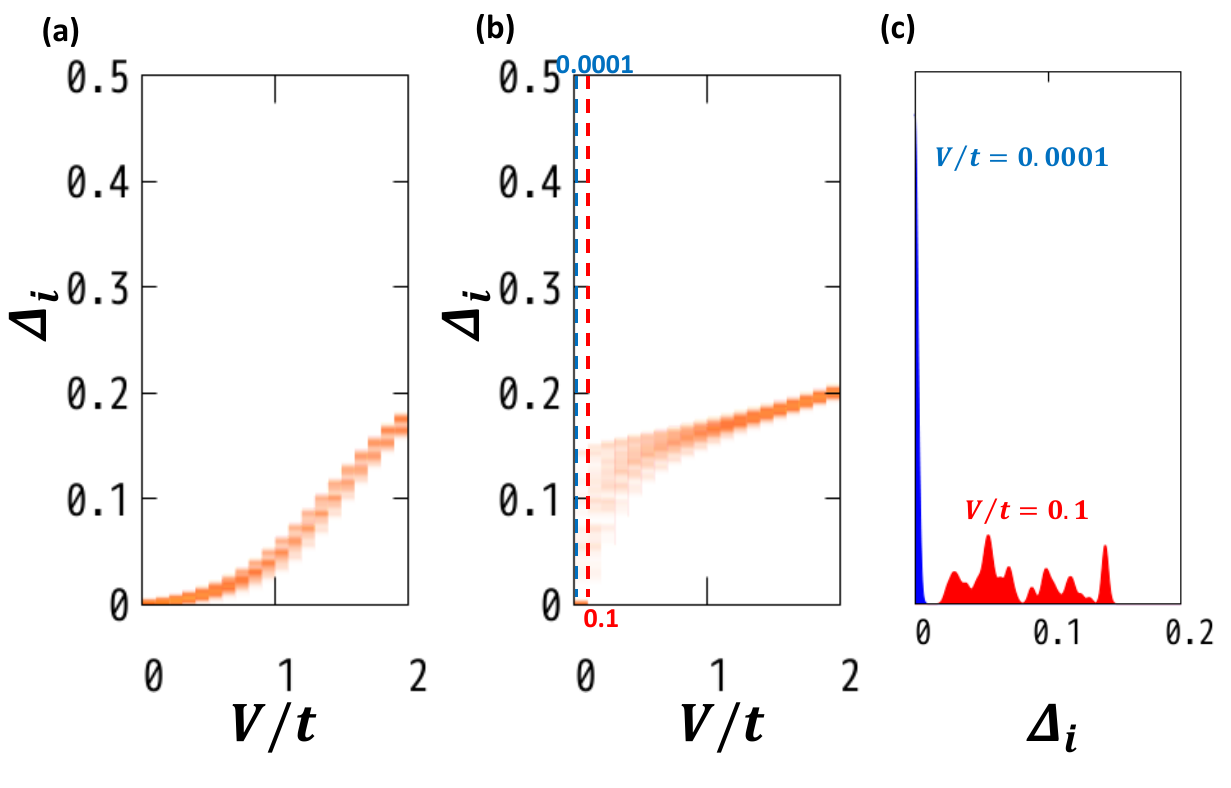}
\caption{
  (Color online)
  Density plot of EI order parameters only five lattice sites around S5 vertices
  for (a)$D/t=-4$ and (b)$D/t\simeq-4.7$ in the center model with $N=10640$. (c) shows the cross section of (b) for indicated $V/t$.
}
\label{S5}
\end{figure}

To make this point clear,
we also show the density plot of the EI order parameter for the five lattice sites
around the S5 vertices in Fig.~\ref{S5}. 
It is found that in the case with $D/t=-4$,
the order parameters slowly increases, which means that,
increasing interband interactions,
the sites are not so active.
Then, the site dependence little appears and
the density plot is almost described by the single curve.
On the other hand, in the case with $D/t\simeq-4.7$,
an interesting site-dependence appears around the small $V$ region.
This means that the sites should be classified into some groups
reflected by the extended structures.
  When $V/t\ll 0.1$, the order parameter is almost zero at each site [see Fig. \ref{S5}(c)]
  since the ground-state degeneracy is not proprotional to the system size
  in the noninteracting limit.

%%%%%%%%%%%%%%%%%%%%%%%%%%%%%%%%%%%%%%%%%%
%%%%%%%%%%%%%%%%%%%%%%%%%%%%%%%%%%%%%%%%%%

%%%%%%%%%%%%%%%%%%%%%%%%%%%%%%%%%%%%%%%%%%%%%%%%%
\begin{figure*}[t]
\centering
\includegraphics[width=\linewidth]{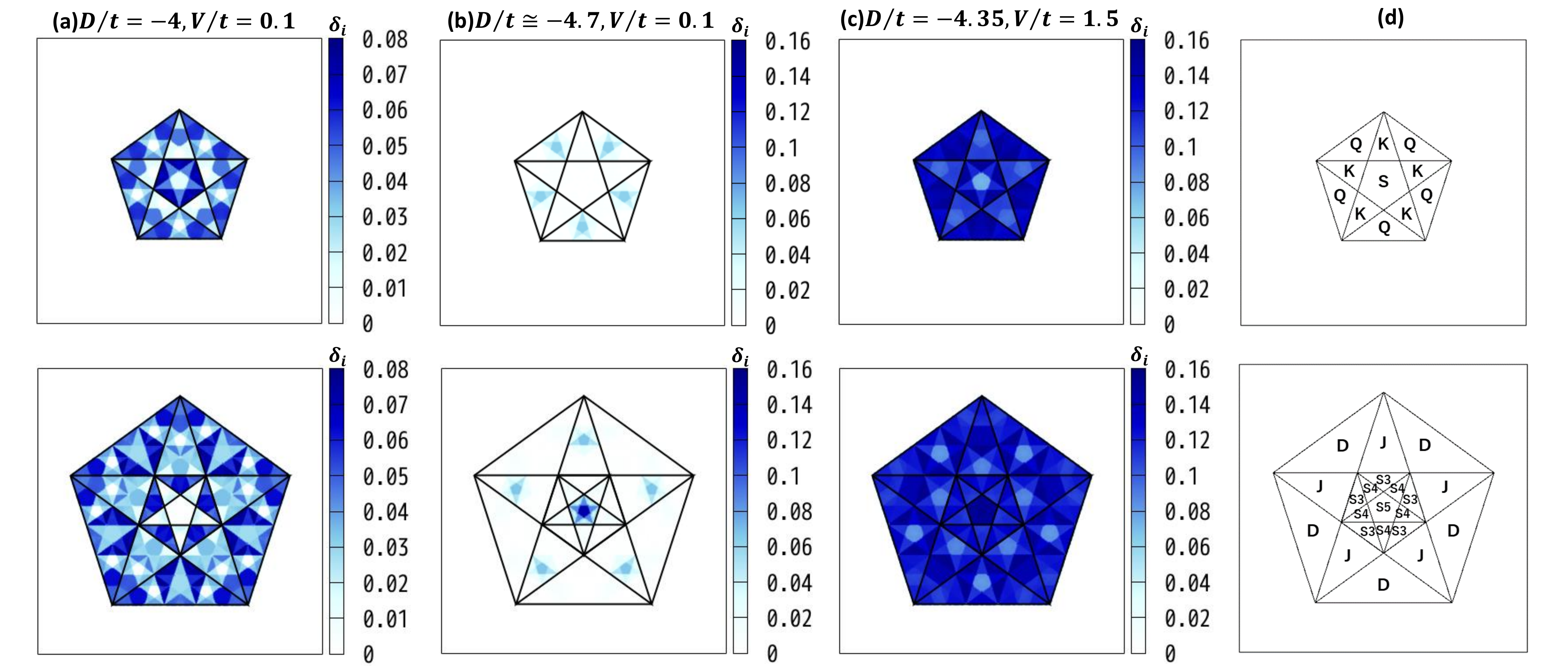}
\caption{
  (Color online)
  The profile of the locally-averaged EI order parameters, $\delta_i=\sum_k \Delta_k/z_i$,
  in the perpendicular space $(\tilde{x}, \tilde{y})$ for the two-band system on the center model
  with $N=28270$ when (a) $(D/t, V/t)=(-4,0.1)$, (b) $(D/t, V/t)\simeq(-4.7,0.1)$, and (c) $(D/t, V/t)=(-4.35,1.5)$.
  Upper (lower) panels are in the $\tilde{z}=0,3$ ($\tilde{z}=1,2$) plane.
  (d) each subdomain is the region for one of the eight kinds of vertices
  shown in Fig.~\ref{VMandCM}(a).
}
\label{perp}
\end{figure*}
%%%%%%%%%%%%%%%%%%%%%%%%%%%%%%%%%%%%%%%%%%%%%%%%
We discuss how large spatial structures are realized in the EI order parameters.
It is useful to analyze the profile in the perpendicular space~\cite{perpendicularspace}.
Each vertex in the Penrose tiling is represented by a five dimensional lattice point
${\bf n} = (n_1,n_2,n_3,n_4,n_5)$ labeled with integers $n_{\mu}$, as shown in Fig.~\ref{VMandCM}(a).
Their coordinates are the projections onto the two-dimensional physical space,
\begin{equation}
\label{eq:r}
\bm{r}=(x,y)=({\bf n}\cdot{\bf e}^x,{\bf n}\cdot{\bf e}^y),
%=\sum_{\mu=1}^{5}n_{\mu}e_{\mu}
\end{equation}
where $e^{x}_{\mu} = \cos(\phi\mu +\theta_{0})$, $e^{y}_{\mu} = \sin(\phi\mu +\theta_{0})$
with $\phi = 2\pi/5$ and $\theta_0$ is constant.
The projection onto the three-dimensional perpendicular space is given by
\begin{eqnarray}
\label{eq:r}
\tilde{\bf r}&=&(\tilde{x},\tilde{y})=({\bf n}\cdot\tilde{\bf e}^x, {\bf n}\cdot\tilde{\bf e}^y),\\
\tilde{z}&=&{\bf n}\cdot\tilde{\bf e}^z,
\end{eqnarray}
where $\tilde{e}^{x}_{\mu} = \cos(2\phi\mu)$, $\tilde{e}^{y}_{\mu} = \sin(2\phi\mu)$, and
$\tilde{e}^{z}_{\mu} = 1$.
%It is known that $\tilde{z}$ takes only four values $\{0,1,2,3\}$,
It is known that $\tilde{z}$ takes only four consecutive integers.
In this paper, we defined four values as $\{0,1,2,3\}$.
In each $\tilde{z}$-plane,
the $\tilde{r}$-points densely cover a region of pentagon shape.
An important point is that in this perpendicular-space representation,
vertices with the similar environment map into the same subdomain
in the perpendicular space.
For examples, eight kinds of vertices in the Penrose tiling have subdomains
in the perpendicular space, as shown in Fig.~\ref{perp}(d).
Therefore, it is useful to clarify the spatial distribution of the local quantities
in the Penrose tiling.
We calculate the local average of the EI order parameters $\delta_i=\sum_k \Delta_k/z_i$,
where $k$ runs the lattice sites defined by the rhombuses sharing the $i$th vertex
and $z_i$ is its number.
Using $\delta_i$ on the vertex sites,
we obtain the EI profile in the perpendicular space.
Figure~\ref{perp} shows the EI order parameter profiles
for $(D/t,V/t)=(-4,0.1)$ and $(D/t,V/t)\simeq(-4.7,0.1)$.
When $D/t=-4$, the detailed patterns appear in the perpendicular space,
as shown in Fig.~\ref{perp}(a).
This suggests that the spatial structure in the EI order parameters
is not classified by the kind of the vertices,
but by a fairly large structure in the Penrose tiling.
This originates from the spatial distribution of the macroscopically degenerate states~\cite{PhysRevB.37.2797}.
On the other hand, different behavior shows up in the case with $D/t\simeq-4.7$,
where finite amplitudes appears only around some regions.
Large amplitudes appear in the S5 domain [see Fig.~\ref{perp}(b)].
In addition, we find in the S5 pentagon domain, 
the star-like structure and
largest $\delta_i$ appears in its center pentagon.
This means that the S5 vertices are further classified into, at least,
three groups depending on the local environments
and the largest EI order parameter appears around the S5 vertices in the cluster-1.
This EI property is in contrast to that for the $D/t=-4$ case
since there are no order parameters around the S5 vertices.
When $(D/t,V/t)=(-4.35,1.5)$, the system shows the EI state with $\bar{\Delta}\simeq0.12$.
In the case, the EI order parameter appears at each site and
its site dependence becomes smaller than the above cases.
Then we have observed the crossover in the EI state,
where the spatial distribution in its order parameters is gradually changed.

We have examined spatial distribution in the EI order parameters.
In these small-interacting regions, the spatial distribution of
the EI order parameters strongly depend on the difference between the energy levels of bands ($D$). 
Namely, the sharp peak structure in the noninteracting DOS yields interesting spatial distribution.
%of the order parameters in the weak-coupling region.
On the other hand, with increasing the interactions, the EI ordered state appears
in the whole lattice sites, where site dependence in the EI order parameters smears.
Therefore, we can say that
the EI states with distinct structure discussed above are adiabatically connected
to each other.

Before summarize the paper, we would like to comment on the relation between the EI phase studied here and the superconductivity studied in previous works~\cite{PhysRevB.95.024509,PhysRevResearch.1.022002}.
As is shown in Appendix \ref{sec:sc_ei}, the attractive Hubbard model studied previously is equivalent to a spinless two-band model with the interband local interaction.
Within the MF theory, the spinless two-band model and the two-band Hubbard model considered in the main text  become equivalent except for the Hartree term, and they become exactly equivalent at $U=V$.
Remember that we focused on the case of $U=2V$, where the Hartree terms can be neglected.
In other words, one can systematically tune the ratio between $U$ and $V$ to discuss the contribution of the Hartree term, which would be an interesting future work.

\section{SUMMARY}
\label{sec:sec5}
In this study, we have examined the excitonic insulating phase in the two-band Hubbard model
on the Penrose tiling.
We have determined the ground state phase diagram of the Hubbard Hamiltonian 
on the vertex and center models, 
and have revealed that the EI phase emerges even with a small interband interaction
at specific values of $D/t$.
This phenomenon is associated with the existence of (nearly) degenerate states,
which is characteristic of the Penrose tiling.
In particular, in the center model, we have found the stable EI phase
due to the exactly degenerate states (confined states and string states) and the nearly degenerate states.
Then we have studied the spatial pattern of the order parameters around these parameter regimes
and have found that it strongly depends on the parameters reflecting different nature between
the exactly degenerate states and the nearly degenerate states.

In this study, we have focused on static features of the EI ordered phases
in the quasiperiodic systems. 
It is an interesting direction to study dynamical aspects of ordered phases
in the quasiperiodic systems,
where one can expect anomalous properties of collective modes such as the phase mode and the amplitude (Higgs) mode in 
the SC phase \cite{Anderson1963,Littlewood1982,Matsunaga2013,Matsunaga2014,shimano2020} and the EI phase \cite{Zenker2014,Murakami2017}.
It is also instructive to consider whether or not
the exciton-Mott transitions~\cite{PhysRevLett.118.067401} occurs in the quasiperiodic system.
These are now under consideration.

\begin{acknowledgments}
  We would like to thank H. Tsunetsugu for fruitful discussions.
  This work was supported by JST CREST Grant No. JPMJCR1901 (Y.M.) and
  Grant-in-Aid for Scientific Research from
  JSPS, KAKENHI Grant Nos. JP19K23425 (Y.M.),
  JP19H05821, JP18K04678, JP17K05536 (A.K.).
\end{acknowledgments}

\appendix

\section{Relevance to Superconductivity}\label{sec:sc_ei}
In this section, we explain the relation between the two-band Hubbard model and
the attractive Hubbard model studied previously to clarify the relation 
between the EI phase and the SC phase.
%Firstly, an exciton is a pair of an electron and a hole and in the EI phase the pair is condensed as the Cooper pair is condensed in the SC phase. 
%This can be seen by applying a particle-hole transformation to the attractive Hubbard model on the Penrose lattice\cite{PhysRevResearch.1.022002} 
%to obtain  the  two-band spinless model on the Perose lattice with the local interband interaction.
%Applying particle-hole transformation to the Hamiltonian of superconductivity, we obtain the  two-band spinless model on the Perose lattice with the local interband interaction.
The Hamiltonian of the attractive Hubbard model~\cite{PhysRevResearch.1.022002} is given by
\begin{align}
\hat{H}^{\rm (SC)}&=\sum_{\langle i,j \rangle \sigma}(-t\hat{a}_{i\sigma}^{\dagger}\hat{a}_{j\sigma}) -\mu_{\rm SC}\sum_{i\sigma}\hat{n}_{i\sigma} +U_{\rm SC}\sum_{i}\hat{n}_{i\uparrow}\hat{n}_{i\downarrow},  \label{schamiltonian}
\end{align}
where $\hat{a}^\dagger$ is a creation operator of the electron and
$\hat{n}_{i\sigma}=\hat{a}^\dagger_{i\sigma}\hat{a}_{i\sigma}$.
$\mu_{SC}$ and $U_{SC}(<0)$ are the chemical potential and the attractive interaction.
Now, we consider the following transformation;
$\hat{a}_{i\uparrow}^{\dagger}=\hat{b}_{i\alpha}^{\dagger}, \hat{a}_{i\downarrow}^{\dagger}=\hat{b}_{i\beta}.$
Here, $\alpha$ and $\beta$ are new indices. 
%With this transformation Eq.~\eqref{schamiltonian},
Then, we obtain 
\begin{align}
\hat{H}^{\rm (SC)}&=\sum_{\langle i.j \rangle}(t\hat{b}_{i\beta}^{\dagger}\hat{b}_{j\beta}-t\hat{b}_{i\alpha}^{\dagger}\hat{b}_{j\alpha})+\frac{U_{\rm SC}}{2}\sum_{i}(\hat{n}_{i\alpha}+\hat{n}_{i\beta})\notag\\
&+\bigl(\frac{U_{\rm SC}}{2}-\mu_{\rm SC}\bigl)\sum_{i}(\hat{n}_{i\alpha}-\hat{n}_{i\beta}) -U_{\rm SC}\sum_{i}\hat{n}_{i\alpha}\hat{n}_{i\beta},
\label{changeschamiltonian}
\end{align}
where we have neglected a constant term.
This Hamiltonian is equivalent to the two-band spinless model with the local interband interaction,
\begin{align}
\hat{H}^{\rm (EXC)}&=\sum_{\langle i,j \rangle}(t\hat{f}_{i}^{\dagger}\hat{f}_{j}-t\hat{c}_{i}^{\dagger}\hat{c}_{j})+\frac{D_{\rm EXC}}{2}\sum_{i}(\hat{n}_{ci}-\hat{n}_{fi}) \notag\\
&-\mu_{\rm EXC}\sum_{i}(\hat{n}_{ci}+\hat{n}_{fi}) +V_{\rm EXC}\sum_{i}\hat{n}_{ci}\hat{n}_{fi}.
\label{excitonhamiltonian}
\end{align}
In Table.~\ref{table:summary}, we summarize the correspondence of the parameters
between the two Hamiltonians Eqs.~(\ref{changeschamiltonian}) and (\ref{excitonhamiltonian}).

\begin{table}[b]
  \begin{tabular}{|c|c|c|} \hline
     parameter & spinless model & attractive Hubbard \\ \hline \hline
    energy splitting & $D_{EXC}$ & $U_{SC}-2\mu_{SC}$ \\ 
    repulsive interaction & $V_{EXC}$ & $-U_{SC}$ \\
    chemical potential & $\mu_{EXC}$ & $-\frac{U_{SC}}{2}$ \\
    \hline
  \end{tabular}
\caption{The correspondence between the two-band spinless model Eq.~\eqref{excitonhamiltonian} and the attractive Hubbard model Eq.~\eqref{schamiltonian}.}
\label{table:summary}
\end{table}

%Now we explain the relation between the spin-less two-band model and the two-band Hubbard model within the MF treatment.
When we apply the MF theory to the spinless model Eq.~\eqref{excitonhamiltonian}
at half filling $(\mu_{\rm EXC}=V_{\rm EXC}/2)$, we obtain
\begin{align}
\hat{H}^{\rm (EXC)}_{\rm MF}&=\sum_{\langle i,j \rangle}(t\hat{f}_{i}^{\dagger}\hat{f}_{j}-t\hat{c}_{i}^{\dagger}\hat{c}_{j})+\frac{D_{\rm EXC}}{2}\sum_{i}(\hat{n}_{ci}-\hat{n}_{fi}) \notag\\
&+V_{\rm EXC}\sum_{i}[(n_{fi}-\frac{1}{2})\hat{n}_{ci} +(n_{ci}-\frac{1}{2})\hat{n}_{fi}] \\
&-V_{\rm EXC}\sum_{i}(\Delta_{i}\hat{f}_{i}^{\dagger}\hat{c}_{i}+h.c.). \notag
\label{excitonhamiltonian}
\end{align}
This is the same as the MF Hamiltonian Eq.~\eqref{eq:MF} at $U=V$ except for the spin degrees of freedom.
On the other hand, the MF Hamiltonian Eq.~\eqref{eq:MF} at $U=2V$ corresponds
to the MF Hamiltonian of the spinless model without the Hartree term.

\bibliography{refs}
\bibliographystyle{jpsj}

\end{document}